\documentclass[%
reprint,
 amsmath,amssymb,
 aps,
prb,
a4paper
]{revtex4-2}

\usepackage{graphicx}
\usepackage{dcolumn}
\usepackage{bm}

\begin{document}

\preprint{APS/123-QED}

\title{Uniaxial strain,  topological band singularities and pairing symmetry changes in superconductors }

\author{Macauley Curtis$^1$}
 \email{mac.curtis@bristol.ac.uk}
 \author{Martin Gradhand$^{1,2}$}%
\author{James F. Annett$^1$}%
\affiliation{1 H. H. Wills Physics Laboratory, University of Bristol, Tyndall Ave, BS8-1TL, UK}
\affiliation{2 Institute of Physics, Johannes Gutenberg University Mainz, Staudingerweg 7, 55128 Mainz, Germany}

\date{\today}

\begin{abstract}
Uniaxial strain affects pairing symmetry states in superconductors by changing the lattice symmetry, 
and by altering Fermi surface topology. Here, we
present a systematic study of these effects within a one-band negative-U Hubbard model for $s$, $p$ and $d$-wave pairing states. We consider a general 2D model that can be applied to superconductors under uniaxial strain, modelled via hopping anisotropy, on a square lattice. The results presented here model an in plane compression along the x-axis, which reduces the lattice symmetry from a tetragonal to an orthorhombic crystal space group. The effects of hopping anisotropy on the different types of gap pairings are explored. We show that changes in $T_c$ are tunable with hopping anisotropy and depend on the orientation of 
the gap function in relation to the opening of the Fermi surface during the Lifshitz transition. In comparing the model 
results to experimental data for the case of Sr$_2$RuO$_4$ it is found that both the $d+id$ and $d+ig$ pairings best describe the 
changes in T$_c$ for the superconducting state in regards to its response to small uniaxial strain.

\end{abstract}

\maketitle

\section{\label{sec:level1} Introduction}

Unconventional superconductivity is defined by a pairing state which spontaneously breaks both gauge 
and crystallographic space group or spin rotational symmetry. In this context the effects of uniaxial strain
can provide a unique probe of the pairing state symmetry, principally by altering the 
crystallographic space group, for example from hexagonal or tetragonal to orthorhombic, in a controlled and
repeatable way.  Group theory, combined with Ginzburg-Landau symmetry arguments can provide a 
clear picture of the 
 effect of strain on different group representations ~\cite{Volovik_Split_Transition, Sigrist_group_theory, JamesGroupTheory, UPT3_Split}.  Specifically a pairing state derived from a degenerate irreducible representation of the symmetry group, 
 will split into two non-degenerate irreducible representations as a result of the symmetry change induced by uniaxial strain. 
 This would then be observable by a splitting of the specific heat jump at T$c$ into a double transition. Furthermore, generic arguments about the coupling of the strain to the order parameter would imply that the splitting in T$_c$ will, for small enough strains, be a linear function of the strain.
 
 Uniaxial strain may also be a useful probe of the pairing state
 even when the pairing state corresponds to a non-degenerate irreducible
 representation. External strain will lead to changes in the physical bond lengths in the crystal, which
 in turn would lead to anisotropy in hoping parameters and hence changes in the Fermi surface shape. These effects
 will be most pronounced at strains where the Fermi surface undergoes a Lifstitz change in topology, accompanied
 by a van Hove Singularity (VHS) in the density of states ~\cite{VanHoveBandfillingTc, JamesVanHoveDoping, Friedel_1989, Markiewicz_1997}. The increase
 in density of states at the VHS would normally be expected to lead to a significant peak in T$_c$ as a function of
 strain, which would be observed for both conventional $s$-wave or non-degenerate $d$-wave or $s^{\pm}$-type 
 pairing symmetries.

The original motivation for the theoretical analysis of strain on unconventional superconductors came in the 1980s
in the context of heavy fermion systems, such as UPt$_3$ in which a split transition at T$_c$ was observed~\cite{Volovik_Split_Transition, UPT3_Split}.  However, more recently the superconductor Sr$_2$RuO$_4$
has been the focus of attention, because of continued controversies about whether or not it is a chiral
$p$-wave triplet superconductor~\cite{Sr2RuO4Rev, KnightShiftPustogow}. Specifically a detailed set of uniaxial strain experiments \cite{HicksQuadraticStrain, SteppkeStrongTc,ResistivityVHS} did not show the expected linear splitting of T$_c$ which would have
been required by the degeneracy of the chiral $p$-wave state. Nevertheless a strong dependence of 
T$_c$ on uniaxial strain was found, with T$_c$ rising from $1.5$K to about $3.5$K at a strain of $\approx -0.5$\% 
\cite{SteppkeStrongTc,Tc_plot_not_hicks, ResistivityVHS}. 
These experiments have contributed to a growing debate about the \textit{exact} nature of the superconducting state in Sr$_2$RuO$_4$~\cite{contreras2019anisotropic, UltrasoundEviTwo, HicksQuadraticStrain,SteppkeStrongTc, ThermoEvi, AlexPetsch, GrinenkoSplitTRSB, Kivelson_2020,Suh_2020,sophie2022}.

The first evidence that Sr$_2$RuO$_4$ is unconventional was provided by experiments showing the strong suppression of the superconducting state by the introduction of non-magnetic impurities \cite{Sr2RuO4Rev}. This is predicted by Anderson's theorem when applied to non s-wave superconductors \cite{AndersonTheorum}.
More specific evidence that Sr$_2$RuO$_4$ was a chiral triplet $p$-wave superconductor was provided by 
the lack of change in spin susceptibility reported in Refs.~ ~\cite{TRSBSCinSrRuO4, PolarizedNeutron}. However more recent measurements of the $c$-axis spin susceptibility show a substantial change which does not support the predictions from the chiral triplet state
~\cite{KnightShiftPustogow, AlexPetsch}.  
Independent support for a two component order parameter with degeneracy was provided by the observation of 
time-reversal-symmetry breaking (TRSB) at T$_c$ ~\cite{TRSBSCinSrRuO4, KerrEffect}.  However the strain experiments gave some indication
that under strain the transition temperature for TRSB might not be identical to T$_c$ \cite{GrinenkoSplitTRSB} at least in strained samples.
This has therefore led to a wide debate about possible pairing states linked to evidence for a two component order parameter model ~\cite{contreras2019anisotropic, ThermoEvi,GrinenkoSplitTRSB, UltrasoundEviTwo}, either with or without degeneracy
at T$_c$ in unstrained samples.

The aim of this paper is to examine the effects of uniaxial strain in a generic microscopic model of
unconventional superconductivity. The original symmetry arguments were made  via Ginzburg-Landau theory \cite{JamesGroupTheory} which leads to some general predictions on the behaviour of the order parameter, but are generally not specific
enough to compare with experimental results. Ginzburg-Landau theory can identify
the leading symmetry breaking terms, linear in strain, but cannot predict non-linear higher-order changes. Hence a microscopic model is needed for comparison providing more detail.
 
Below we examine in detail the effect of strain in a simple negative $U$ model of superconductivity on a square lattice.
We treat strain as a small uniaxial perturbation introduced into the hopping parameters which changes the tetragonal lattice to an orthorhombic lattice breaking the x-y symmetry. For simplicity we consider a one band model. The band parameters
are chosen to be similar to those of  the $\gamma$-band of Sr$_2$RuO$_4$ to allow some limited comparison to experiments. 
But this is not intended to be a fully accurate model of the full band structure of Sr$_2$RuO$_4$ under strain. 

In the following sections we first define our model of the strain dependent band structure,  $\varepsilon_k$. We then examine changes in $T_c$ and the gap function corresponding to a variety of different superconducting pairing states, including
$s$-wave, chiral $p$-wave, $d_{x^2-y^2}$, $s^{\pm}$, $d+id$ and $d+ig$.  
In cases where the order parameter degeneracy is split by the uniaxial strain we examine the splitting in the self-consistent
order parameter in the two channels. As well as changes to the order parameter symmetry, 
strain can also have a significant impact on the transition temperature, $T_c$. There are several possible mechanisms but the simplest is  the fact that strain can move the system through a Lifshitz transition in the Fermi surface topology accompanied by a van Hove singularity (VHS) in the density of states at the Fermi level. Other strain related effects on $T_c$ could also include softening of certain phonons related to crystal lattice instabilities, or changes to Fermi surface nesting, the spin-fluctuation spectrum, or moving a system away from or closer to a Mott instability by changing the effective electronic bandwidth. These latter effects are beyond the scope of this paper, where we concentrate on the effects of the Lifshitz transition assuming a constant effective pairing interaction.

\section{Theory}

\subsection{\label{sec:level2} Microscopic Theory}

Here, we present a simple model using a one band negative-U Hubbard model ~\cite{HubbardModel, Hubbard_model_book}, applied to a 2D square or rectangular lattice  - upon which we allow electrons to sit on the lattice sites and hop between nearest and next nearest neighbours. 
We take the standard form for the Hamiltonian\cite{JRobbingThesis}
\begin{equation}
H = H^0 + V^{(1)} ,
\label{eq:Hubbard_Hamiltonian}
\end{equation}
where $H^0$ is the kinetic term with hopping between electron sites on the lattice accounted for via the hopping integrals $t_{{\bf r},{\bf r}^{'}}$ ~\cite{BDG_book, PaulMillerThesis, JRobbingThesis}
\begin{equation}
    H^0 = \sum_{{\bf r},{\bf r}^{'}, \sigma}  t_{{\bf r}, {\bf r}^{'}}~c^{\dagger}_{{\bf r} \sigma}c_{{\bf r}^{'}\sigma}
    ~ \text{.}
\label{eq:H0_Realspace}    
\end{equation}

On the square lattice we  set $t_{{\bf r}, {\bf r}}=\varepsilon_0 $, $t_{{\bf r},{\bf r}^{'}}=-t$  for nearest neighbours and 
$t_{{\bf r},{\bf r}^{'}}=-t'$ for next nearest neighbours. Then uniaxial strain along the $x$ direction modifies the hopping
to $t_x$ for nearest neighbour bonds along $x$ and $t_y$ for nearest neighbour bonds along $y$. The four next neighbour bond lengths remain equivalent so we retain a single hopping parameter, $-t'$.  Transforming to $k$-space we have the Hamiltonian

\begin{eqnarray}
    H^0 &=&   \sum_{{\bf k}\sigma} \Bigg(\varepsilon_0 -2\big( t_x cos(k_x a) + t_y cos(k_y b)\big) \nonumber \\
    & & ~~~~~~~ - 4t^{'}\big(cos(k_x a) cos(k_y b)\big) \Bigg) c^{\dagger}_{{\bf k} \sigma} c_{{\bf k} \sigma} \nonumber \\ 
   & =&  \sum_{{\bf k} \sigma} \varepsilon_{\bf k} c^{\dagger}_{{\bf k} \sigma} c_{{\bf k} \sigma},
\label{H0_full}    
\end{eqnarray}
where $a$ and $b$ are the lattice constants of the rectangular unit cell and  the single particle energy band dispersion is $\varepsilon_{\bf k}$. 
Results are presented in units of the unstrained hopping parameters, t$_0$ which is evaluated at t$_0$ $=$ 81.62meV.

We assume the general interaction between two particles at positions ${\bf r}$ and ${\bf r}^{'}$, $V^{(1)}$, is given by a generalized Hubbard type interaction
\begin{equation}
    V^{(1)} = \frac{1}{2} \sum_{{\bf r}, {\bf r}^{'} \sigma \sigma^{'}}V({\bf r}-{\bf r}^{'}) c^\dagger_{{\bf r} \sigma} c_{{\bf r} \sigma} c^{\dagger}_{{\bf r}^{'} \sigma^{'}} c_{{\bf r}^{'} \sigma^{'}}\ \text{,}
\label{eq:V1_Realspace}    
\end{equation}
where the creation and annihilation operators ($c^{\dagger}_{{\bf r}\sigma}$ and $c_{{\bf r}\sigma}$) take their usual meaning \cite{JamesBook}.  
Transforming to $k$-space the components of the effective pairing interaction, 
given Cooper pairs with zero centre of mass momentum can be written as  
\begin{equation}
    V^{(1)}=  \sum_{{\bf k},{\bf k}^{'}} V_{{\bf k}, {\bf k}^{'}}
    c^{\dagger}_{{\bf k} \sigma}  c^{\dagger}_{-{\bf k} \sigma^{'}} c_{-{\bf k}^{'} \sigma} c_{{\bf k}^{'} \sigma^{'}}\ \text{,}
\label{eq:V1_k_space}   
\end{equation}
 where 
\begin{equation}
    V_{{\bf k}, {\bf k}^{'}} = \sum_{r,r'} V({\bf r}-{\bf r}^{'}) e^{i ({\bf k} - {\bf k}^{'}) \cdot ({\bf r}-{\bf r}^{'})}\ \text{.}    
\label{eq:Vkk_prime_space}   
\end{equation}
  
 In order to analyse different Cooper pair symmetries it is useful to redefine the pairing interaction
 into symmetry distinct pairing channels.  To do this
 we treat $V_{{\bf k}, {\bf k}^{'}}$ as a linear operator which we define via an eigenvalue equation
\begin{equation}
    \sum_{{\bf k}^{'}} V_{{\bf k},{\bf k}^{'}} \Gamma_i({\bf k}^{'})   =  U_i \Gamma_i({\bf k}) \ \text{,}
\label{eq:v_kk_prime_eigenvalue}   
\end{equation}
 where $U_i$ is a eigenvalue corresponding to eigenvector $\Gamma_i({\bf k})$.
The properties of the group representations will ensure that the eigenvectors $\Gamma_i({\bf k})$ 
can be classified according to the different irreducible group representations of the appropriate symmetry group, in this case either square, or rectangular. The corresponding eigenvalue, $U_i$, is the effective pairing strength in that
symmetry channel. 
We use the linear expansion of an operator to rewrite $V_{{\bf k},{\bf k}^{'}}$ as a sum of the basis eigenvectors 
\begin{equation}
    V_{{\bf k}, {\bf k}^{'}} = \sum_i U_i \Gamma_i({\bf k}) \Gamma_i({\bf k}^{'})\ \text{.}  
\label{eq:linear_expansion_of_an_operator}     
\end{equation} 
Here, we used the fact that the eigenfunctions $\Gamma_i({\bf k})$ can  be chosen as real valued functions for the cases of the $D_{4h}$ and $D_{2h}$ point groups, and assumed that the orthogonal eigenfunctions are normalized to obey
$ \sum_{\bf k}  \Gamma_i({\bf k})^2  = 1 $.
In principle we have an infinite set of eigenvectors for a general operator, but assuming our original
pairing interaction $V({\bf r},{\bf r}^{'})$ was limited to either on-site, nearest neighbour or next nearest neighbor
bonds, we typically have only a unique eigenfunction or pair of degenerate eigenfunctions for each symmetry channel, as specified in the relevant character tables given in Appendix 1. 
 
 We solve the k-dependent BCS gap equation~\cite{BDG_book, SupMetalsandAlloys, schrieffer1999theory}.  Using the eigenvector expansion, Eq. \ref{eq:linear_expansion_of_an_operator}, of the pairing interaction
 the gap equation   can be written in the form
  \begin{equation}
     \Delta_i =  \frac{ U_i}{2} \left( \sum_{\bf k} 
      \Gamma_i({\bf k})  
     \frac{\Delta_{\bf k}}{E_{\bf k}} (1 - 2f(E_{\bf k}))  \right) ,
     \label{eqn:GapEquation}
 \end{equation} 
where the k-dependent gap function is 
 \begin{equation}
     \Delta_{\bf k} = \sum_i \Delta_i \Gamma_i({\bf k}) .
 \end{equation}
and the $\Delta_i$ are the order parameter amplitudes in each pairing channel~\cite{alexandrov2003theory,Tinkham}.
 Here, as usual,  $E_{\bf k} = \sqrt{\varepsilon_{\bf k}^2 + |\Delta_{\bf k}|^2}$ and $f(E_{\bf k})$ is the Fermi-Dirac function evaluated at eigenenergy $E_{\bf k}$. 
 Note that the factor of $\frac{1}{2}$ from Eq.~(\ref{eq:V1_Realspace}) has been absorbed into the constant $U_i$.
  In the case of 
 a non-degenerate irreducible representation this gives simply $\Delta_{\bf k} = \Delta \Gamma({\bf k})$ where $\Delta$ is the numerical value of the gap function.
 Furthermore, in the case of conventional local BCS singlet $s$-wave pairing 
 $V_{{\bf k},{\bf k}^{'}}=-U$
 we have $ \Gamma({\bf k})=1 $ and 
 $\Delta_{\bf k} = \Delta$ implying that we recover the usual $s$-wave BCS gap equation \cite{schrieffer1999theory}.    Note that the expression for $E_{\bf k}$ is valid for all 
 equal spin pairings presented here, either spin singlet 
  pairing (for which $\Delta_{\bf k}=\Delta_{-{\bf k}}$) or 
 opposite spin triplet  pairing (with ${\bf d}_{\bf k}=-{\bf d}_{-{\bf k}} = \Delta_{\bf k}\hat{\bf z} $). More general types of triplet pairing, such as nonunitary states, 
 would require a more complex gap equation. 
 The self consistent equation (Eq.~\ref{eqn:GapEquation}) can be solved numerically on a grid in k-space making use of the group tables, table \ref{table:D2h} and \ref{table:D4h}, to choose the corresponding function $\Gamma({\bf k})$ given by table \ref{table:gap}~\cite{JamesGroupTheory}.

Physical insight into the effects of strain on $T_c$ can be 
gained by considering the linearized gap equation in which a single order parameter
component $\Delta_i$ is non-zero, but infintesmimally small.  Linearizing
Eq. (\ref{eqn:GapEquation})
 we recover the usual BCS expression for $T_c$,
\begin{equation}
    1  = \frac{U}{2} \int_{-\infty}^{\infty} N(\varepsilon) \frac{1}{\varepsilon}  (1 - 2f(\varepsilon)) d\varepsilon\ \text{,}
\label{eqn:GapEquationTc}    
\end{equation} 
but where the effective density of states is weighted
by the eigenvector $\Gamma_i({\bf k})$ of the specific pairing channel 
which becomes non-zero at $T_c$ 
\begin{equation}
    N(\varepsilon) = \sum_{{\bf k}} \Gamma_i({\bf k})^2 \delta(\varepsilon-\varepsilon_{\bf k}) .
\label{eqn:DensityOfStates}    
\end{equation} 
The presence or absence of a van Hove singularity in this weighted density of states at the Fermi energy ($\varepsilon=0$) will then be a useful guide to 
whether or not strain leads to an enhanced $T_c$ in the corresponding pairing channel $\Delta_i$.

 \subsection{\label{sec:level2} Strain induced lattice anisotropy}

The remaining consideration is the inclusion of lattice anisotropy resulting from the uniaxial strain. In our approach it is most convenient to simply model this by adjusting the hopping integrals for the $x$ and $y$ oriented bonds in the crystal lattice, as resulting from the shrinking or increasing the physical distance of the bonds in real space. 
 To better model the physical system we note that the uniaxial strains $\epsilon_{xx}$ and $\epsilon_{yy}$ are not fully independent, but linked by the Poisson ratio. In compressing the $x$-axis an expansion of the y-axis will result from the Poisson ratio of the material (Sr$_2$RuO$_4$: $V_{xy} \approx -0.4$ \cite{HicksQuadraticStrain}). 
 Representing the changes to the square lattice parameter, $a$, as $\delta_x$ and $\delta_y$ respectively, we therefore assume $\delta_y = V_{xy} \delta_x$, and the corresponding changes to the $x$ and $y$ nearest neighbour hopping integrals will be given by
 \begin{equation}
     t_y-t_0= V_{xy} (t_x - t_0) ~ .
     \label{eq:anisotropichopping}
 \end{equation}  
 To avoid additional unknown parameters we assume, for simplicity, that the next neighbor hopping $t'$ is unchanged by strain. We also assume that the pairing interaction
 $V({\bf r}-{\bf r}^{'})$ is unchanged by the strain. The overall strength of the pairing interaction in the unstrained lattice is tuned to give a $T_c$ of about $1.5$K, similar to that of Sr$_2$RuO$_4$, in each channel, $U_i$ considered below. 
 Finally, note that the strain induced changes to the hopping integrals alter the
 band structure $\varepsilon_{\bf k}$ significantly, especially as the van Hove singularity is approached. The changed band energies can lead to an unwanted side effect, changes in the total number of electrons ($N_e$) given by 
 \begin{equation}
    N_e =  2 \sum_{\bf k}  f(\varepsilon_{\bf k})\ \text{.}
\label{NumberOfElectrons}    
\end{equation}
Clearly the total number of electrons has to remain constant as a function of strain, and so we adjust the on-site energy $\varepsilon_0$ together with the hopping integrals $t_x$ and $t_y$  keeping $N_e$ constant and the chemical potential at zero for all values of the strain.

\section{Results}

\subsection{\label{sec:NormalState} Normal State}

\begin{figure}[t]
\includegraphics[width=42mm]{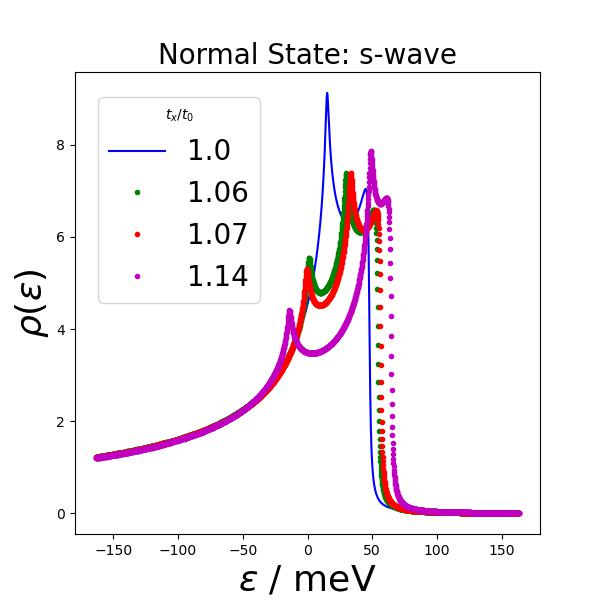}
\includegraphics[width=42mm]{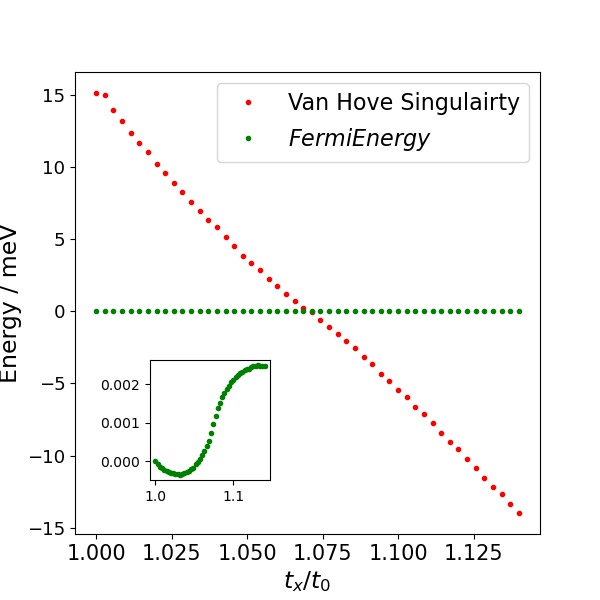} \\ 
\includegraphics[width=42mm]{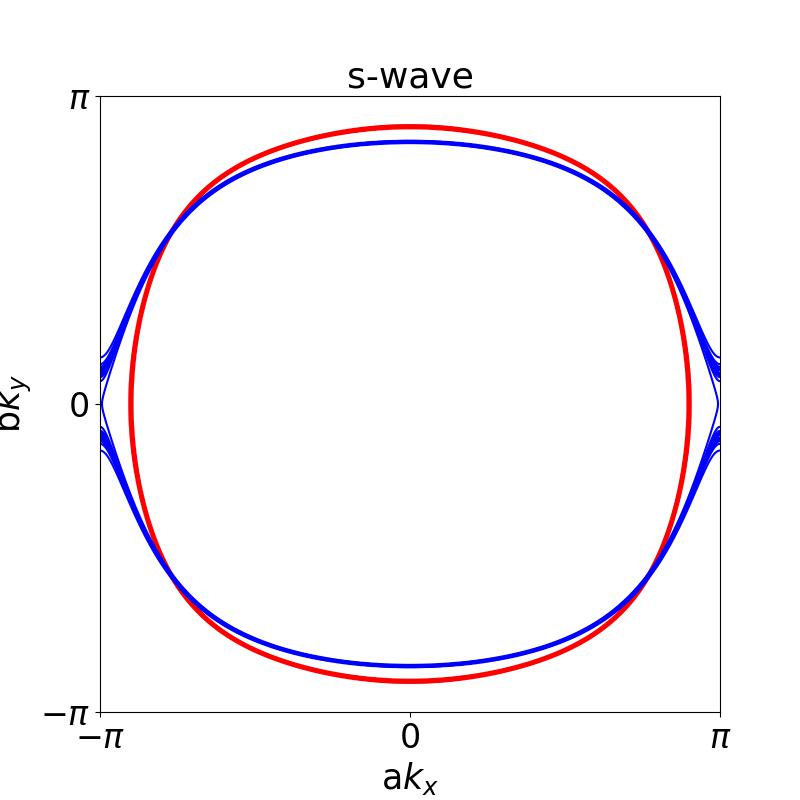}
\includegraphics[width=42mm]{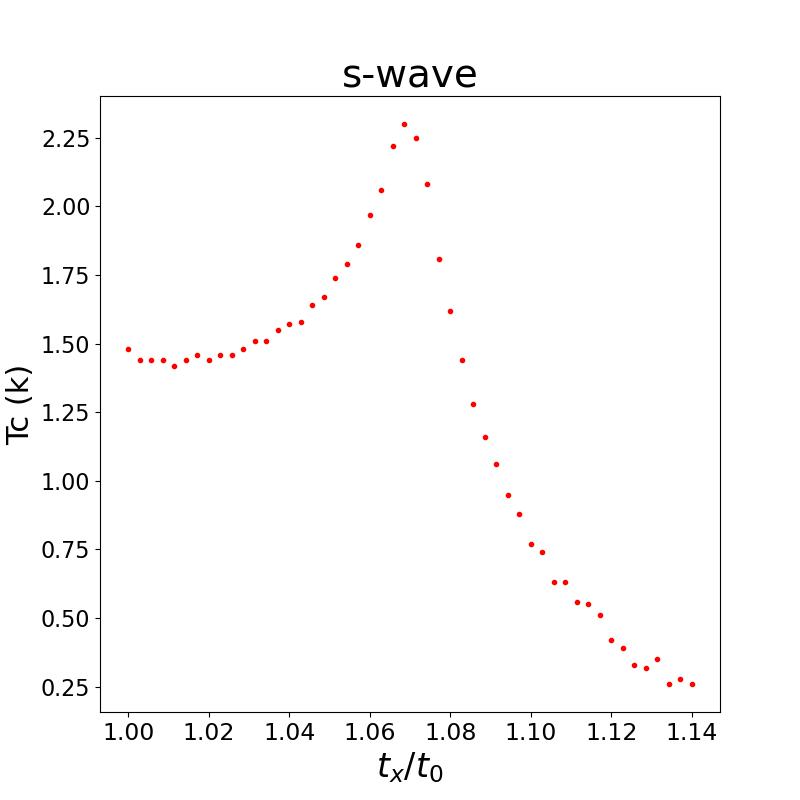}
\caption{\label{fig:top_trans} a) The normal state dos presented at various lattice anisotropies.
b) The energy (red dots) for which the  $VHS_{Lifshitz}$ occurs in the normal state as a function strain $t_x/t_0$. The green dots show the Fermi energy as a function of strain which changes marginally due to the self-consistent procedure defined by Eq.~\ref{NumberOfElectrons}. The crossing of both dependencies defines the strain ($t_x = 1.07 t_0$) for which the $VHS_{Lifshitz}$ crosses the Fermi energy.  
c) The constructed Fermi Surface for an s-wave model at $t_x = t_0$ (red) and $t_x = 1.08 t_0$ (blue).
d) $T_c$ as a function of strain for conventional s-wave pairing. }
\end{figure}

First, we study the density of states (DOS) in the normal state as a function of strain induced lattice anisotropy (see Fig.~\ref{fig:top_trans}~(a)). The unstrained square lattice (blue line)
has two distinct van Hove singularities (VHS), one corresponding to the upper band edge 
and the other associated with the band saddle point in $\varepsilon_{\bf k}$ at ${\bf k}=(\pi/a,0)$ and ${\bf k}=(0,\pi/a)$. As a function of the hopping anisotropy 
the upper band edge singularity shifts only slightly relative to the chemical potential ($\varepsilon=0$),
but a more significant change is that the single VHS of the unstrained lattice splits into two distinct peaks associated with the separate band saddle points at ${\bf k}=(\pi/a,0)$ and ${\bf k}=(0,\pi/b)$, now no longer degenerate. 

For all anisotropies we can identify two VHS, the one at lower energy gradually crossing the Fermi energy at higher tensions and a second one at high energies. The first gives rise to the Lifshitz transition and we will label it $VHS_{Lifshitz}$. For the uncompressed system its energy is at roughly $15$ $meV$ corresponding to the band at ($k_x, k_y$)$=$($\pi, 0$) in good agreement to experiment~ \cite{Fermi_level_value}. The second VHS moves upwards until eventually it merges with  the band edge in our one band model, and this singularity will subsequently be labeled as $VHS_{\varepsilon_{max}}$.

Fig.~\ref{fig:top_trans}~(b)  shows the energy of the lower of these two peaks 
 as it moves down in energy and 
crosses the Fermi level at an anisotropy of about $t_x = 1.07 t_0$. This point can be identified as the Lifshitz transition in Fermi surface topology, changing from a closed to open
topology, as shown in  Fig.~\ref{fig:top_trans}~(c).

As the Lifshitz transition is a topological change of the Fermi surface which can be caused by applying high pressure \cite{Lifshitz1960ANOMALIESOE} it is important to identify the precise tension for which this transition appears. We summarize this transition in Fig.~\ref{fig:top_trans}~(b), where $VHS_{Lifshitz}$ crosses the Fermi energy at about $t_x = 1.07 t_0$ inducing the Lifshitz transition. This is further underlined by Fig.~\ref{fig:top_trans}~(c) 
indicating the opening of the spherical Fermi surface into the neighbouring Brillouin zone for a tension of $t_x = 1.07 t_0$ just above the Lifshitz transition. Strictly speaking, what we visualise in Fig.~\ref{fig:top_trans}~(c) is a contour plot of Eq.~\ref{eqn:GapEquation} for the case of an s-wave $\Gamma(k)=1$ state. This way we are able to visualize the regions of the BZ, which predominantly contribute to the superconducting gap. For an s-wave state on a spherical Fermi surface this procedure essentially traces the Fermi surface. In the following we will see how for different pairing symmetries distinct regions of the Brillouin zone will contribute.

\begin{figure*}[t]
\includegraphics[width=34mm]{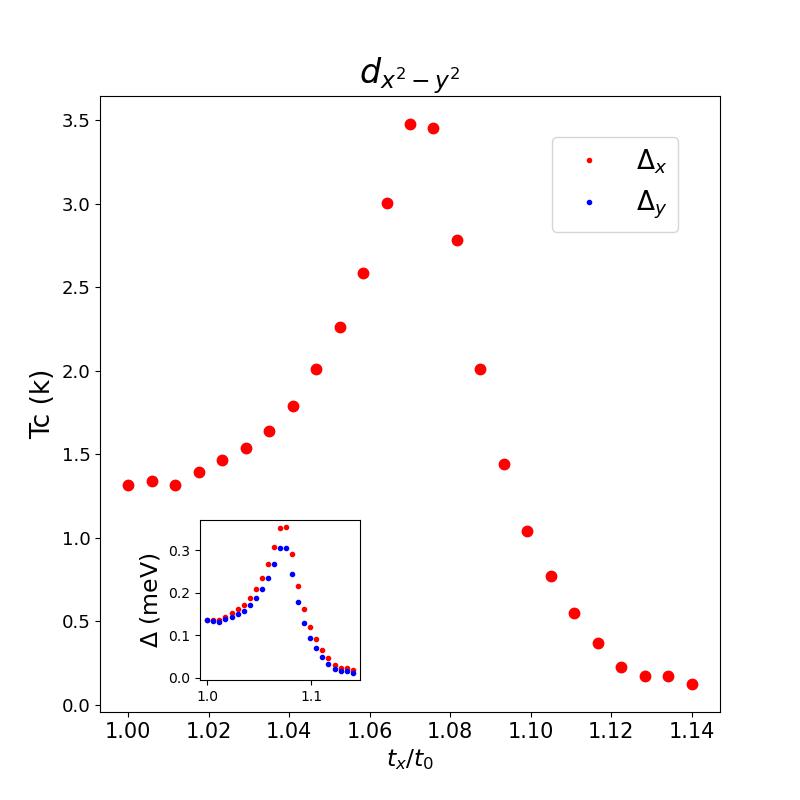}
\includegraphics[width=34mm]{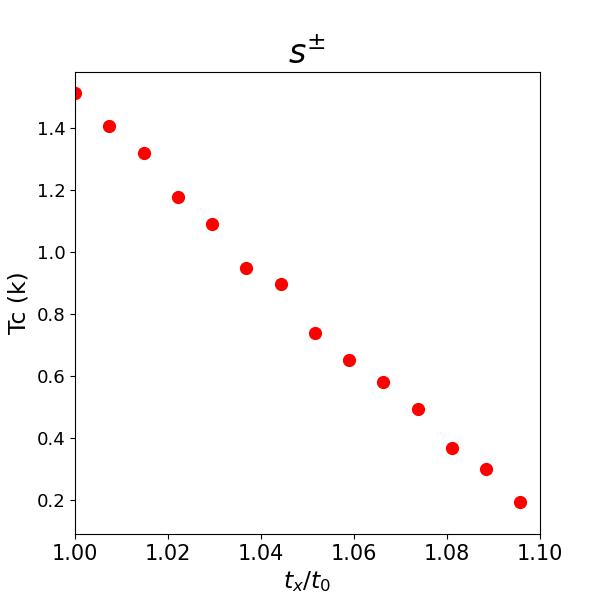}
\includegraphics[width=34mm]{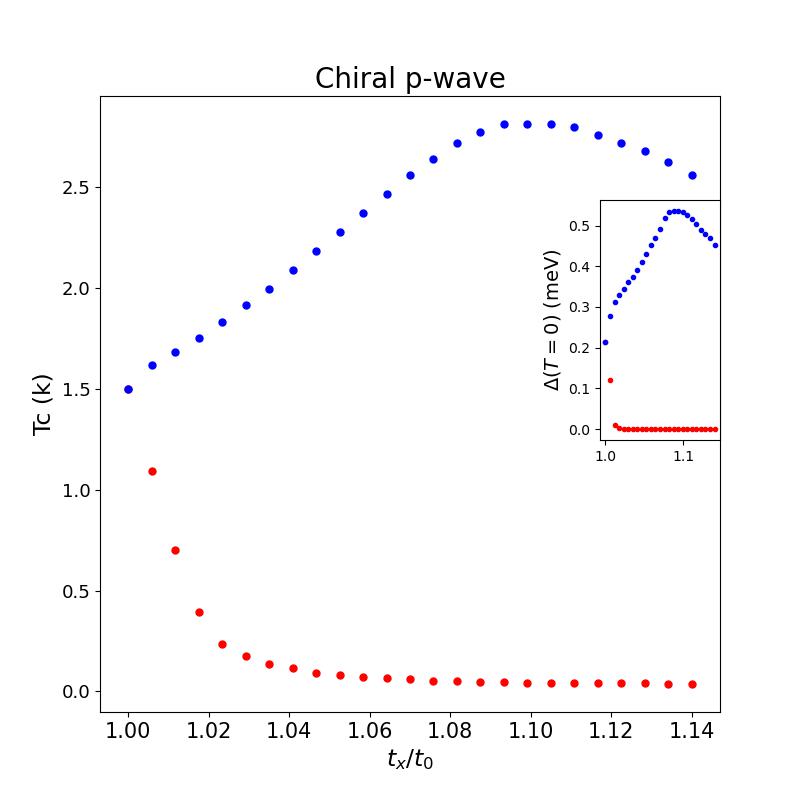}
\includegraphics[width=34mm]{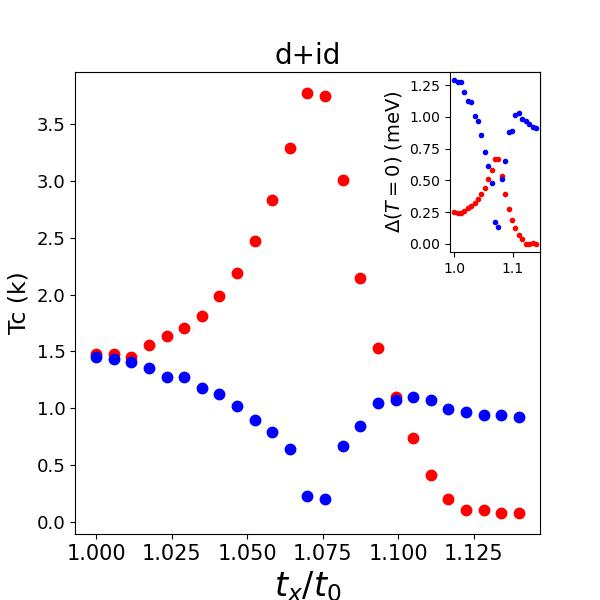}
\includegraphics[width=34mm]{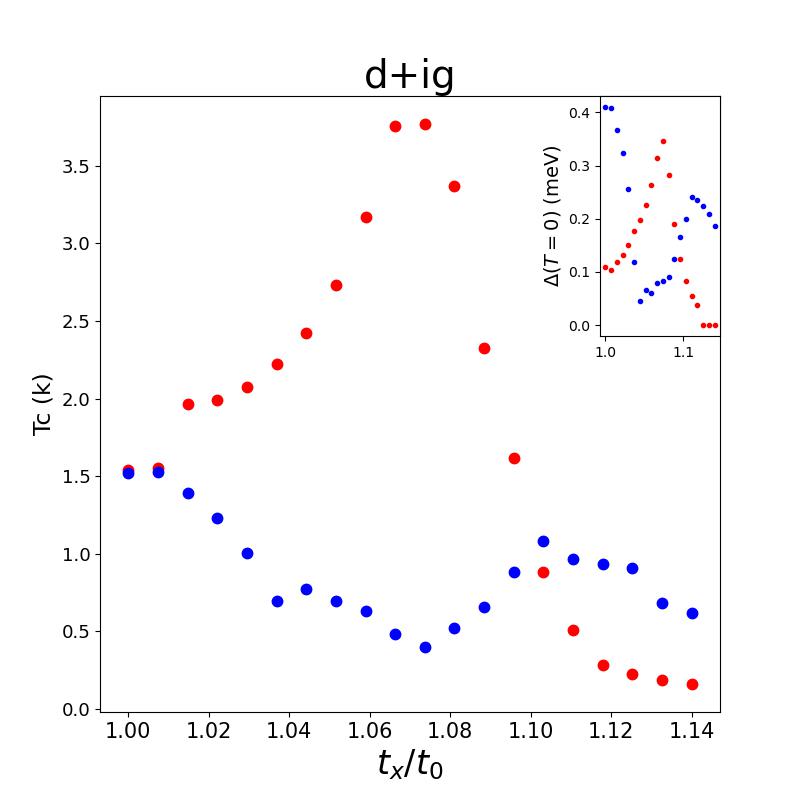}
\caption{\label{fig:Tc_all} $T_c$ as a function of hopping anisotropy 
modelling uniaxial strain for 
a) d-wave ($d_{x^2-y^2}$) pairing - 
$\Gamma_{\bf k} =  \cos{(k_xa)} - \cos{(k_yb)} $,
b) extended s-wave pairing $\Gamma_{\bf k} =  \cos{(k_xa)} +  \cos{(k_yb)} $, (c)
chiral p-wave pairing - $\Gamma_x({\bf k})  = \sin{(k_xa)} $, $\Gamma_y({\bf k})= \sin{(k_yb)} $.
d) $d+id$ pairing - $\Gamma_1({\bf k}) =  \cos{(k_xa)} - \cos{(k_yb)}$,
$\Gamma_2({\bf k}) = \sin{(k_xa)}\sin{(k_yb)}  $ and,
e) $d+ig$ pairing - $\Gamma_1({\bf k}) =  \cos{(k_xa)} - \cos{(k_yb)}$, 
$\Gamma_2({\bf k}) = ( \cos{(k_xa)} - \cos{(k_yb)} )  \sin{(k_xa)}\sin{(k_yb)} $, (omitting normalization constants for clarity). 
Insets to (a) (d) and (e) are of the value of the zero temperature gap parameters in the two component calculations as a function of the anisotropy. For the pure d-wave, the inset shows a two component system modelling pure d-wave in the x (red) and y(blue) channels. For the $d+id$ and $d+ig$, the channels are color coded such that the $d_{x_2-y_2}$ channels are red, the the $d_{xy}$ and $g$ channels are blue in both the main figure and the insets.
} 
\end{figure*}

\subsection{Superconducting $T_c$ and pairing symmetry}

In a first step we present the critical temperature $T_c$ as a function of hopping anisotropy for conventional s-wave pairing in Fig.~\ref{fig:top_trans}~(d). In this case the 
BCS equation for $T_c$, Eq. \ref{eqn:DensityOfStates}, applies with the weighting function 
$\Gamma({\bf k})=1$. So $T_c$ is determined by the full dos and therefore has a clear maximum at the  point when the VHS crosses the Fermi energy, as seen in  Fig.~\ref{fig:top_trans}~(b).  The increase in $T_c$ near the peak has a maximum at about 2.25K, which is substantial, but not as large as observed in Sr$_2$RuO$_4$ ~\cite{SteppkeStrongTc}.
This discrepancy is to be expected since Sr$_2$RuO$_4$ is not a conventional BCS s-wave superconductor~\cite{Sr2RuO4Rev}.

Now we turn to consider the changes in $T_c$ for all unconventional gap symmetries  considered here (see Fig.~\ref{fig:Tc_all}). We see that the d-wave models,
$d_{x^2-y^2}$, $d+id$ and $d+ig$ give the largest increases in $T_c$, as seen in 
Figs. ~\ref{fig:Tc_all}~(a), (d) and (e), respectively.   
 For the conventional BCS s-wave pairing, T$_c$ is enhanced by about $50\%$ relative to the zero temperature. In contrast, the enhancement is about $150\%$ for the two d-wave symmetry models considered.   
 We can see qualitatively different behavior  for the case of extended $s$-wave ($s^\pm$), where $T_c$ decreases smoothly with lattice anisotropy and there is no peak in $T_c$ at any value of lattice anisotropy. 
 The case of chiral $p$-wave pairing, shown in Fig.~\ref{fig:Tc_all}~(c) is also 
 qualitatively different.  Importantly, and discussed above already, the uniaxial anisotropy breaks the symmetry between the p$_x$ and p$_y$ pairing leading to an enhancement of T$_c$ for $p_y$ pairing and a suppression of T$_c$ for the $p_x$ pairing. 
  Here, T$_c$ is initially changing only linearly with the changing hopping anisotropy. The linear splitting of the two degenerate $T_c$'s  is expected on symmetry grounds~\cite{JamesGroupTheory} (also see Appendix 1), although Ginzburg-Landau symmetry arguments alone cannot explain the unequal slopes of the two $T_c$'s.    In marked contrast, for the $d+id$ and $d+ig$ pairing states there is also a splitting of  $T_c$ in two channels degenerate at zero strain
 but in this case the splitting is quadratic, rather than linear, in strain. The T$_c$ increase in the y channel does not match the expected maximum found experimentally.
 The difference arises because in the $p_x+ip_y$ case the zero strain degeneracy is required by symmetry, while in the $d+id$ and $d+ig$ cases the degeneracy is accidental  (a point 
 discussed further in  Appendix 1 below).  In principle the symmetry breaking anisotropy
 also mixes the $d_{x^2-y^2}$ and $s^\pm$ states but, as we show in the inset to Fig.~\ref{fig:Tc_all}(a), the difference between $\Delta_x$ and $\Delta_y$ (where $\Delta_{\bf k}=\Delta_x \cos{(k_x a)}- \Delta_y \cos{(k_y b)}$) is negligible even near the Lifshitz point, and so we can continue to regard the pairing state as having a dominant $d_{x^2-y^2}$ symmetry.

\begin{figure*}[t]
\includegraphics[width=42mm]{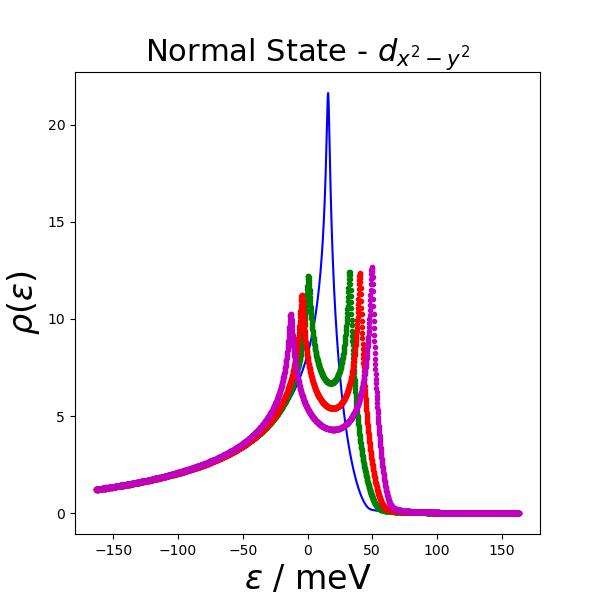}
\includegraphics[width=42mm]{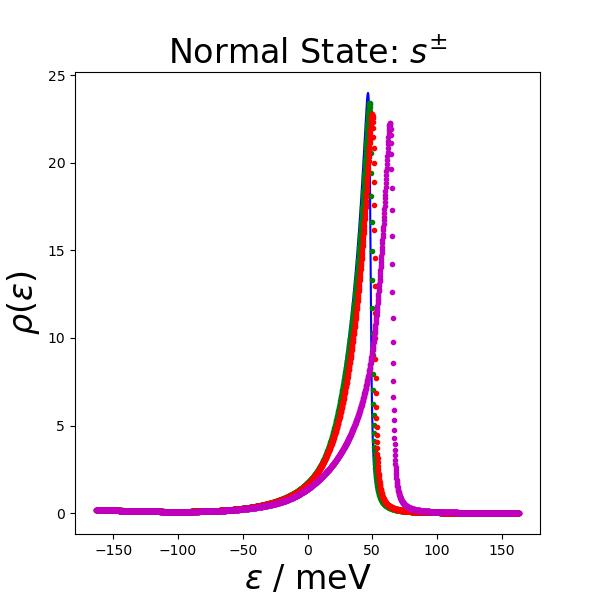}
\includegraphics[width=42mm]{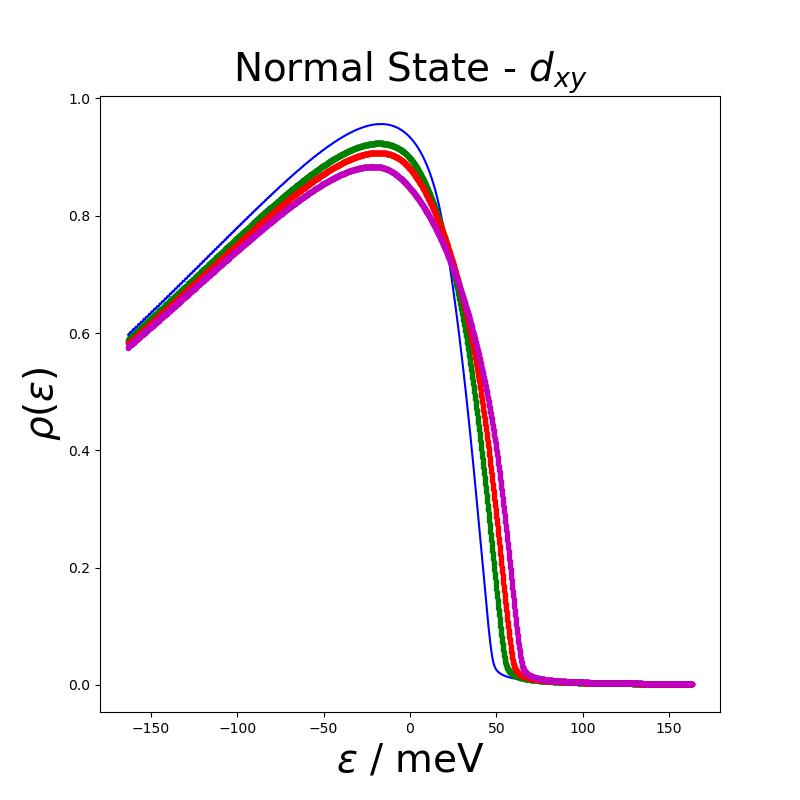}\\
\includegraphics[width=42mm]{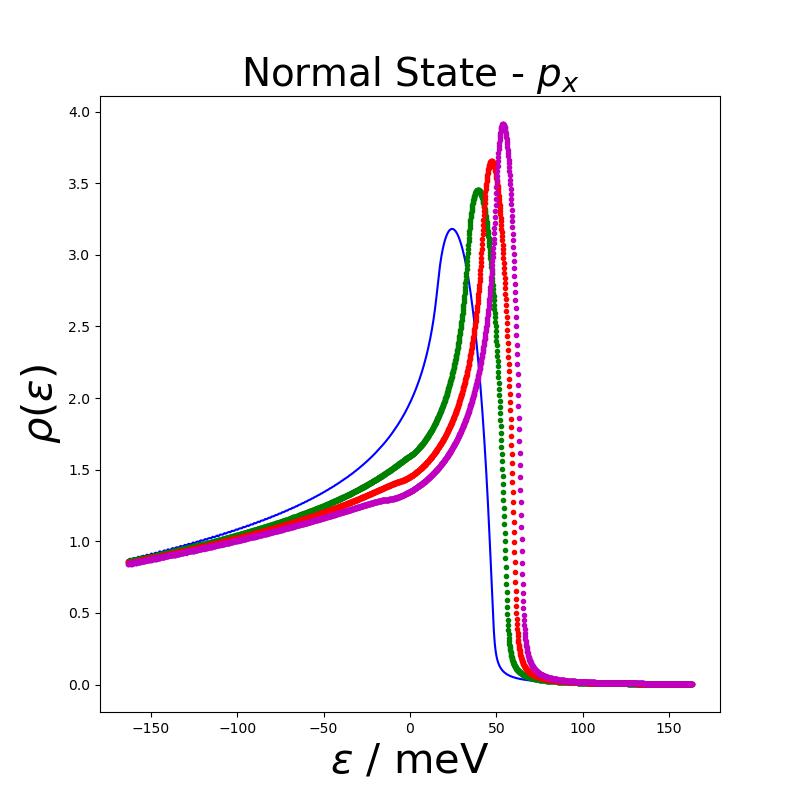}
\includegraphics[width=42mm]{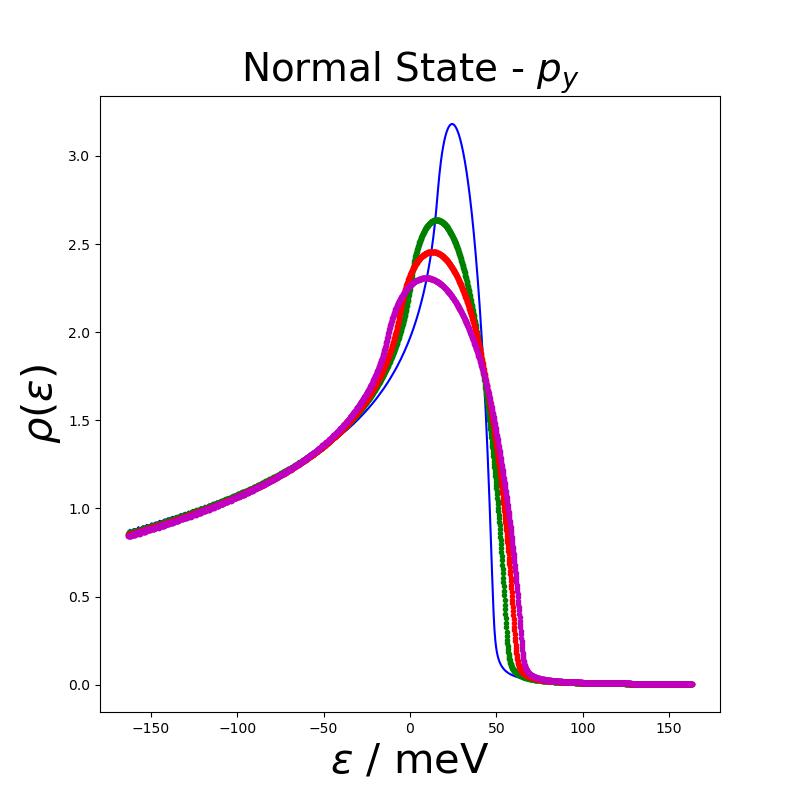}
\includegraphics[width=42mm]{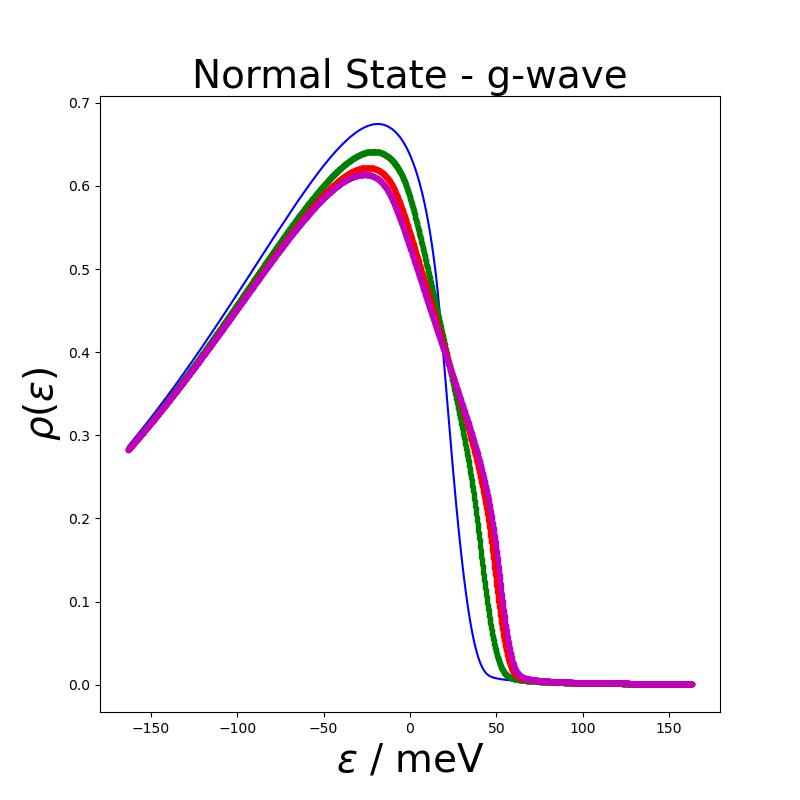}
\caption{\label{fig:dos-all} Normal state density of states at various anisotropies for a) d-wave ($d_{x^-y^2}$), b) d-wave ($d_{xy}$), c) chiral p-wave ($p_x$), d) chiral p-wave ($p_y$) and e) g-wave pairings. The hopping anisotropies shown are the same as given by Fig.~\ref{fig:top_trans}~(a).
}
\end{figure*}

\begin{figure*}[t]
\includegraphics[width=42mm]{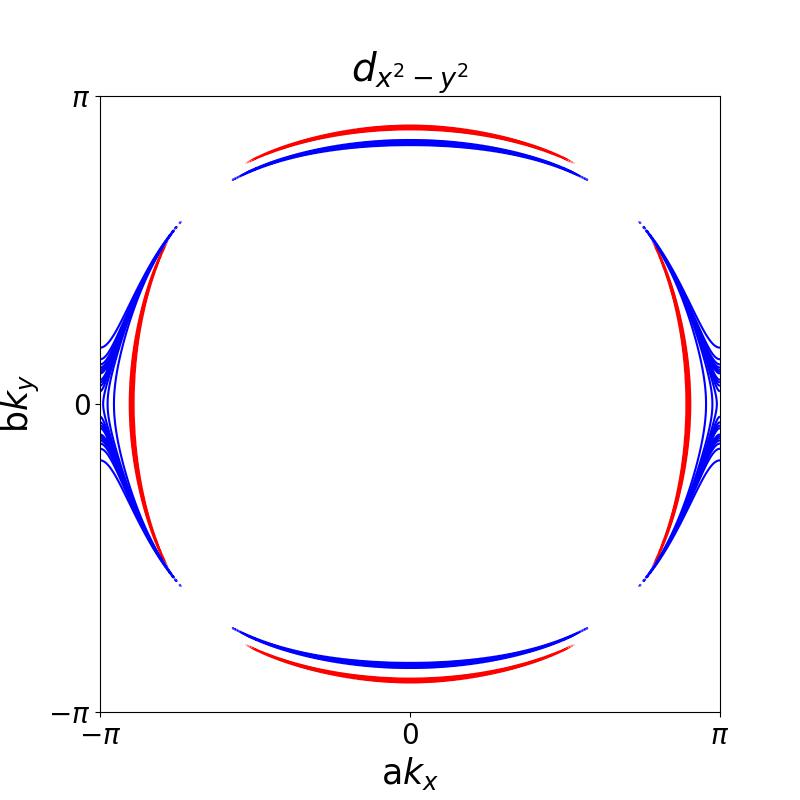}
\includegraphics[width=42mm]{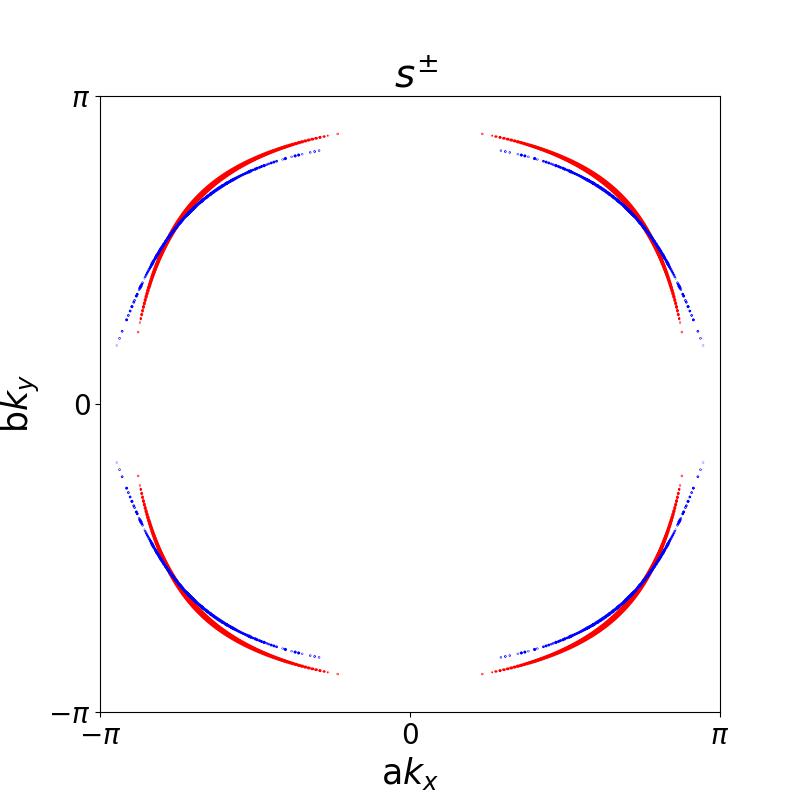}
\includegraphics[width=42mm]{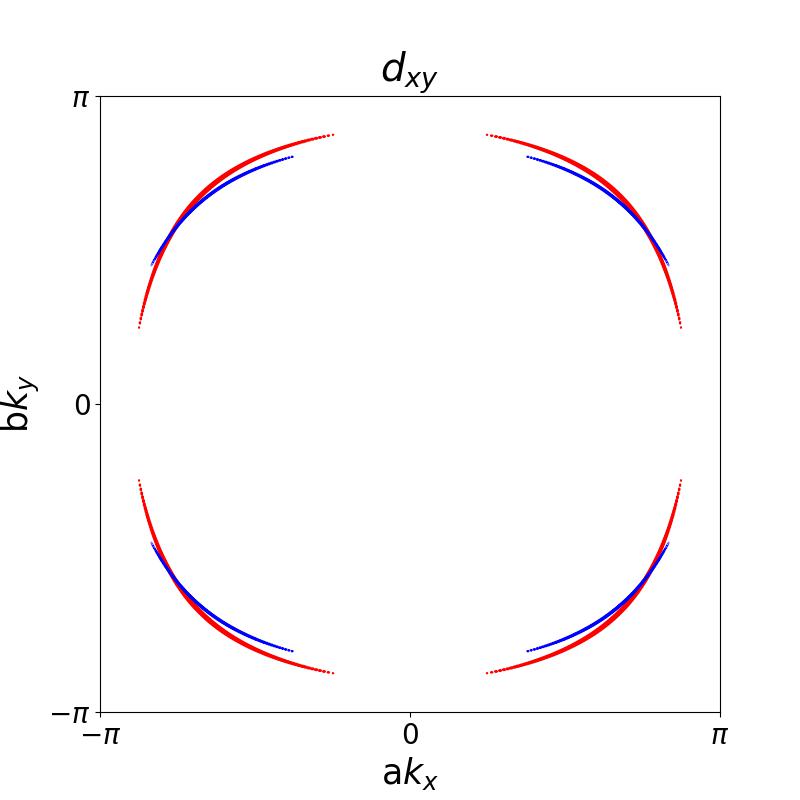}\\
\includegraphics[width=42mm]{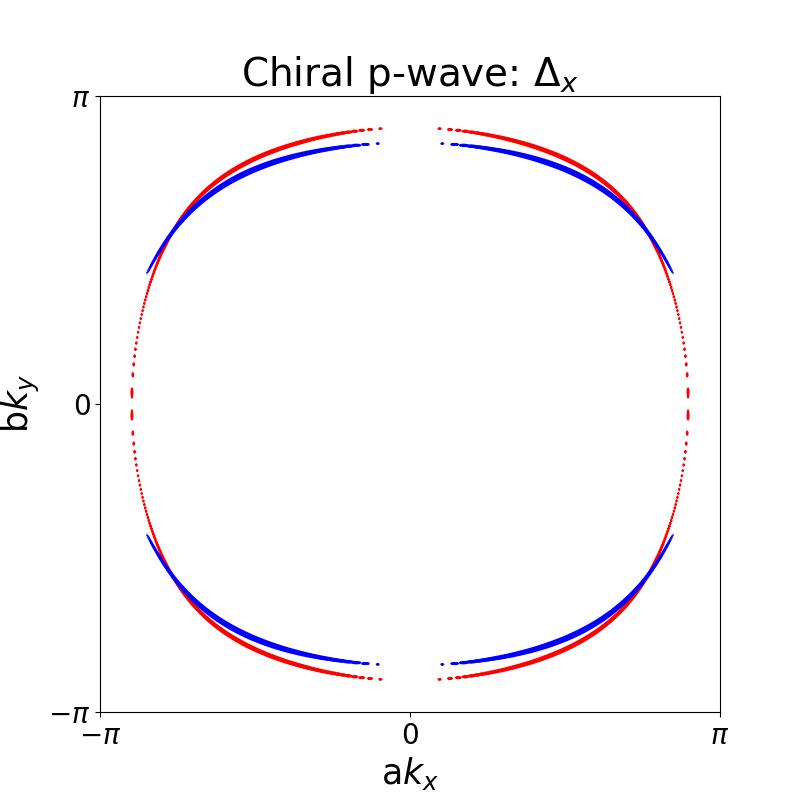}
\includegraphics[width=42mm]{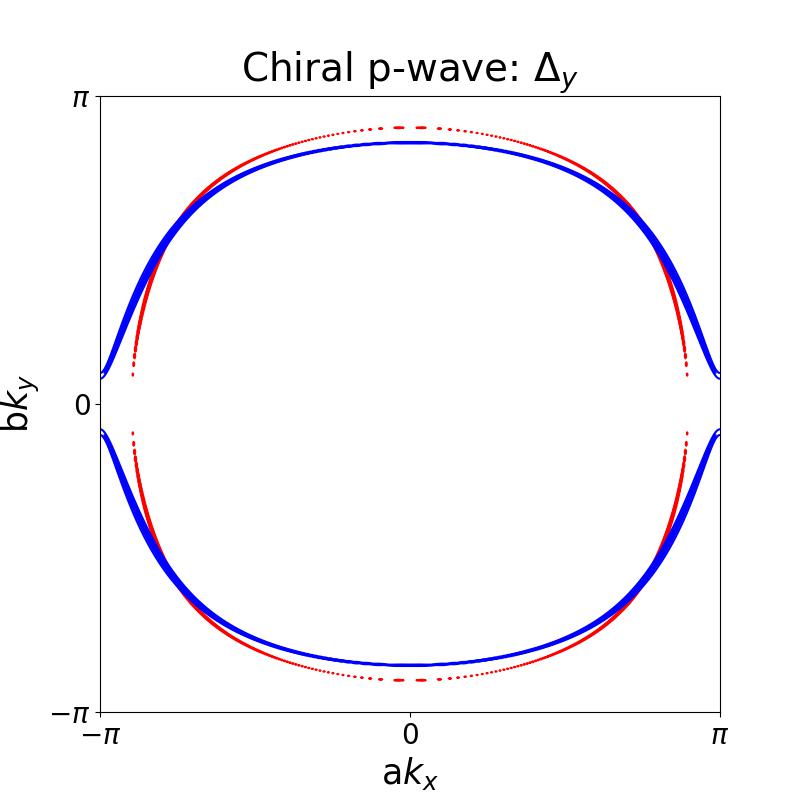}
\includegraphics[width=42mm]{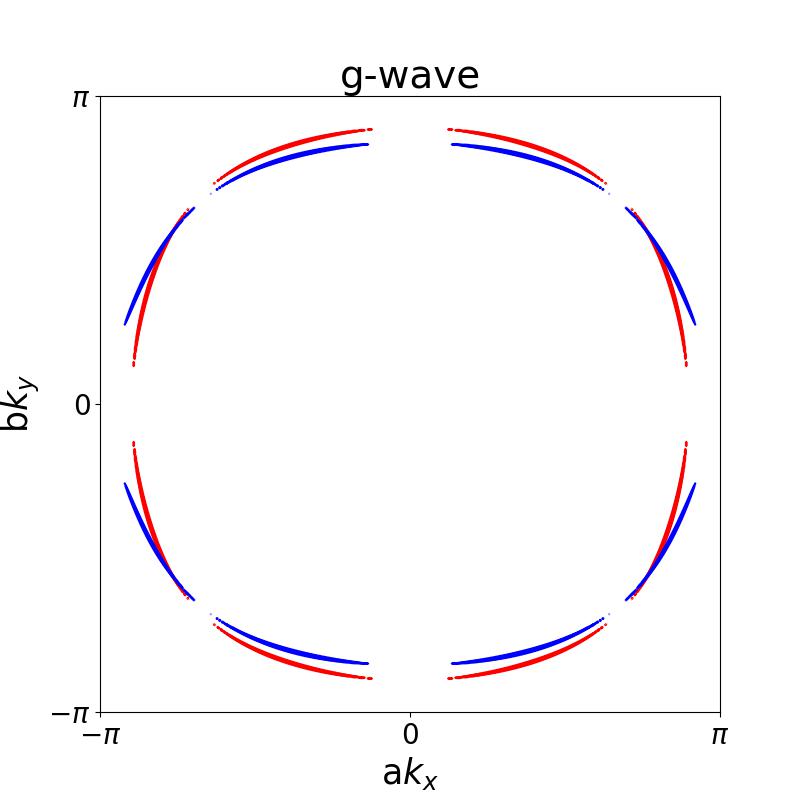}
\caption{\label{fig:Fermi-all} A snapshot of the change in the constructed Fermi surface - calculated from the  k-space points that contribute to the self consistent integration of $\Delta$ from Eq.(\ref{eqn:GapEquation}) plotted at $t_x = t_0$ (red) and $t_x = 1.08 t_0$ (blue) for a) d-wave ($d_{x^-y^2}$), b) d-wave ($d_{xy}$), c) chiral p-wave ($p_x$), d) chiral p-wave ($p_y$) pairings) and e) g-wave. The red contour is the unstrained Fermi surface and the blue contour is the Fermi surface strained past the Lifshitz transition at $\frac{t_x}{t_0} = 1.08$. Coefficients a and b represent the atomic bond length in real space presented here to keep a square Brillouin zone. Unstrained, as expected, a=b. As soon as hopping anisotropy is applied this no longer holds and they are distinctly different.}
\end{figure*}

 This different behavior is a consequence of the 
 fact that the weighted density of states in the $s^\pm$ channel has no VHS singularity at the Fermi level for any lattice strain, Fig.~\ref{fig:dos-all}~(b). There is no peak in $T_c$  because the weighting function $\Gamma({\bf k})^2$  in this channel is zero at the Lifshitz point.
 
  To obtain more insight in the effects of changing symmetries on the DOS we present 
  in Fig.~\ref{fig:dos-all} the weighted dos (Eq.~(\ref{eqn:DensityOfStates})) for various possible gap symmetries. In that framework Fig.~\ref{fig:top_trans}~(a)  represents the corresponding density of states for conventional s-wave symmetry $\Gamma(k)=1$. Among the other symmetries considered only the d-wave ($d_{x^2-y^2}$) symmetry induces two clearly defined VHSs, with one of them crossing the Fermi energy at the Lifshitz point.   In the other cases shown, chiral p$_x$, p$_y$, d${_{xy}}$ and g-wave symmetry, the only visible VHS is pushed to higher energies essentially merging with the VHS arising from the band edge. A similar behavior is seen in the case of extended s-wave symmetry (Fig.~\ref{fig:dos-all}~(b)) where the only VHS present moves up towards the band edge, rather than crossing the Fermi level. 
  Fig.~\ref{fig:dos-all}(d)-(e) shows that the weighted dos for chiral p$_x$ and p$_y$ symmetry are affected strongly but in opposing ways. While the VHS energy moves up with the induced anisotropy for the p$_x$ symmetry, it moves down for the p$_y$. This breaking of the symmetry between the to possible states is a direct result of the breaking of the symmetry via the uniaxial strain we apply. The $d_{xy}$ and g-wave density of states, shown in Figs.~\ref{fig:dos-all}(c) and (f) are relatively unaffected with only minor changes around the Fermi-level resulting from the anisoptropy.
  
  To clarify the origin of these differences in the weighted density of states
  in a microscopic picture we show the regions predominantly contributing to the superconducting
  gap equation
  for all pairing symmetries in Fig.~\ref{fig:Fermi-all}. They are the equivalent to Fig.~\ref{fig:top_trans}~(c) using the respective basis functions for each pairing symmetry. As before the figures are contour plots of the integrand of Eq.~(\ref{eqn:GapEquation}).

  From Fig.~\ref{fig:dos-all}~(a) it is clear that 
  the reason that the $T_c$ for the $d_{x^2-y^2}$ state is enhanced more strongly than for on-site s-wave is the increased weighting of the VHS peak which crosses the Fermi level at the Lifshitz point.  In both the total and d-wave weighted densities of states, Fig.~\ref{fig:top_trans}~(a)
  and Fig.~\ref{fig:dos-all}~(a),
  the single VHS of the unstrained lattice splits into two 
 peaks, one of which eventually crosses the Fermi energy at the Lifshitz transition. 
 For the d-wave case the enhancement in the dos is larger than for conventional s-wave pairing, because
 the weighting function $\Gamma({\bf k})^2$ for $d_{x^2-y^2}$ pairing has a maximum
 in $k$-space at the point ${\bf k}=(\pi/a,0)$. We can see this directly in the 
 corresponding k-space plot for the d-wave case, Fig.~\ref{fig:Fermi-all}, compared
 to the on-site s-wave case Fig.~\ref{fig:top_trans}~(c). 
 
 In contrast, there is no peak in $T_c$ at the Lifshitz point in any other pairing channel
shown in Fig.~\ref{fig:Tc_all} as a direct result of the absence of a VHS crossing the Fermi level
in the respective weighted densities of states in Figs.~\ref{fig:dos-all}~(b)-(f). For example, the extended s-wave pairing case
has no enhancement of the weighted density of states at the Lifshitz point 
(Fig.~\ref{fig:dos-all}~(b)) because the corresponding weighting function $\Gamma({\bf k})$
has nodal lines which pass through the Lifshitz point, Fig.~\ref{fig:Fermi-all}(b).
Similar nodal lines exist for the $p_x$, $p_y$, $d_{xy}$ and $g$ weighting functions, visible in
Fig.~\ref{fig:Fermi-all}(c)-(f).  The splitting in the 
 $T_c$ for $p_x$ and $p_y$ pairing can be understood from the changes in the
 weighted density of states in these pairing channels,
 with one of the two channels having a slightly
 enhanced dos at the Fermi energy.  
  The small linear
 increase in the upper $T_c$ in Fig.~\ref{fig:Tc_all}(c) is consistent with the 
 small enhancement in the dos shown in Fig.~\ref{fig:dos-all}(e).  
 The fact that the lower $T_c$ drops rapidly for even the very small strain values 
 shown in Fig.~\ref{fig:Tc_all}(c)  is a consequence of the non-linearity in the coupled $p_x$, $p_y$ gap equation, Eq.~\ref{eqn:GapEquation}.  This 
 difference can also be seen in the $p_x$ and $p_y$ k-space plots of Figs.~\ref{fig:Fermi-all}(d)-(e).
 
 The much stronger splitting in the $T_c$ values for the $d+id$ case is a consequence
 of the existence of the strong VHS singularity crossing the Fermi level in the $d_{x^2-y^2}$
 pairing channel, which is totally absent in the $d_{xy}$ channel (Figs.~\ref{fig:dos-all}(a) and (c)). Again the nodal points of the $d_{xy}$ pairing function  (Fig.~\ref{fig:Fermi-all}~(c))
 coincide with the Lifshitz point, while the nodal lines of the $d_{x^2-y^2}$ gap are
 far from the Lifshitz point, Fig.~\ref{fig:Fermi-all}(a).  Therefore, $T_c$ 
 strongly increases in the $d_{x^2-y^2}$ channel. The $d_{xy}$ $T_c$ was chosen to be 
 degenerate with the $d_{x^2-y^2}$ channel at zero strain, but that ``accidental'' degeneracy
 is not preserved once the strain is applied. Based on the weighted density of states alone we would not expect such a strong decrease in $T_c$ for the $d_{xy}$ channel, but at temperatures below the upper of the two transition temperatures the non-linearity of the gap equation (Eq.~\ref{eqn:GapEquation}) means that an enhancement in pairing in one channel is accompanied by
 a suppression of the pairing in the other channel. Very similar behavior is found in the $d+ig$ scenario~\cite{Kivelson_2020}, which also assumes an accidental degeneracy present in the unstrained samples, which is no longer the case in the presence of uniaxial strain. Interestingly, for our model parameters
 the $d_{xy}$ and $g$ pairing channels eventually become the higher of the two $T_c$'s for the 
 $d+id$ and $d+ig$ states for strains beyond the VHS topological transition point, Figs.~\ref{fig:Tc_all}(d) and (e) .

\section{Discussion}

We now consider the above results in the context of the recent uniaxial strain experiments on
Sr$_2$RuO$_4$ ~\cite{HicksQuadraticStrain,SteppkeStrongTc,GrinenkoSplitTRSB,Tc_plot_not_hicks,ResistivityVHS}.
Firstly we note that the two dimensional one band Hubbard model has been utilised here due to its simplicity. Clearly a more realistic three dimensional 3-band model gives a more accurate overall picture for Sr$_2$RuO$_4$ ~\cite{ReenaPaper, JamesInterlaying}. Nevertheless, only one of the three Sr$_2$RuO$_4$ Fermi surface sheets, the $\gamma$ sheet, lies sufficiently close to the topological
Lifshitz transition to be tunable through the transition in experiment ~\cite{SteppkeStrongTc,ResistivityVHS}, and so it remains a reasonable assumption that a single band model can capture the relevant physics near to the 
topological transition. The experiments also appear to be consistent with the simpler two-dimensional band model. In a full three dimensional band the logarithmic VHS in the dos is split into two cusp singularities~\cite{VanHove}. The fact that the overall band structure of Sr$_2$RuO$_4$ is highly two dimensional~\cite{Sigrist_group_theory} as well as the fact that experimental strain measurements~\cite{ResistivityVHS} seem consistent with a single Lifshitz point (as in 2d) rather than a pair (as in 3d) implies that the assumption of a two-dimensional band seems well justified.

In the full three band model there is
also a significant spin orbit coupling, SOC, associated with the Ru d-orbitals, which is not included in the one band model ~\cite{James_SO, S_O_Coupling_one}.  SOC is likely to be most significant in the case of spin triplet pairing states, but even in the case of singlet d-wave pairing properties such as the low temperature Knight shift can be strongly influenced by the SOC interaction~\cite{ReenaPaper}.  However the general symmetry principles (discussed below in Appendix 1) relating to the Ginzburg-Landau theory near to $T_c$ are not affected by SOC at all in the case of singlet pairing~\cite{JamesGroupTheory}. For the triplet case we can also omit explicit SOC from the model,  provided that the effective SOC energy scale is stronger than the energy scale for pairing~\cite{James_SO, S_O_Coupling_one}. For example, this will be true if the
SOC simply locks the triplet ${\bf d}-{\bf k}$ vector to a specific orientation, such as $\hat{\bf z}$ as in the chiral $p_x+ip_y$ state examined above.

A further complication in a more realistic multiband model is that it would require changing many unknown parameters in the model corresponding to the change caused by uniaxial hopping anisotropy.
This contrasts with the simplicity of the one band model which  
has the advantage of requiring only one significant parameter which we must take from experiment, 
effectively related to the Poisson ratio of $x$ and $y$ uniaxial strains. 
Under our assumption, the changes in the second neighbor hopping integral $t'$ are negligible compared to the changes in $t_x$ and $t_y$ arising from the uniaxial strain. Then the only
strain related variables become $t_x$ and $t_y$, which are related by the assumption
that both hoppings vary linearly in the corresponding strain and that the ratio
of these variations is given by the Poisson ration, Eq.~\ref{eq:anisotropichopping}.
We present our results in the previous section as functions of the hopping anisotropy $t_x/t_0$, rather than specific physical strain values of Sr$_2$RuO$_4$.  Comparisons to the case of 
Sr$_2$RuO$_4$ then simply need to map from our critical value for the Lifshitz transition ($t_x \approx 1.07 t_0$) to the experimental strain where the transition is observed, as about $\epsilon_{xx} \sim - 0.5 \%$~\cite{SteppkeStrongTc,ResistivityVHS}. With this comparison of scales the 
ranges of lattice strains plotted  in Figs.~\ref{fig:Tc_all}-\ref{fig:dos-all} coincide
with the ranges of compressive lattice strains from $0$ to $-1\%$ studied
experimentally~\cite{HicksQuadraticStrain}.  For the anisotropy dependence of $T_c$ in the case of the extended s-wave pairing Fig.~\ref{fig:Tc_all}~(b) the figure has been cut of at the strain at which the gap function becomes zero.

Above we have laid out the assumptions for treating the potential $V(r,r^{'})$ as a tune-able constant. However it should be noted that this could be better treated including the position dependence since the bond length will change with strain. The strains used here are small and hence we have assumed the strain to have a low impact on the change of the potential. It would, of course, be straightforward to allow the pairing interaction to explicitly depend on the lattice strain, but at the expense of adding another unknown parameter into the model.

With the above caveats aside; we move onto the interesting results obtained by
comparing our results, Fig.~\ref{fig:Tc_all}, with the recent experiments. 
We can confidently say that within the confines of this model, the $d+id$ and $d+ig$ pictures of superconductivity in $Sr_2RuO_4$ match well with experimental data in comparison to the behaviour of other pairings presented. Furthermore, the d-wave picture maps out very well in comparison to the experimental data presented ~\cite{Tc_plot_not_hicks, SteppkeStrongTc, ResistivityVHS} as shown in Figure \ref{fig:Tc_all}~(d)-(e). The conventional one component d-wave model also reproduces the increase in $T_c$ however it shows total suppression of $T_c$ after the Lifshitz transition which is not in agreement with experiment unlike the $d+id$ and $d+ig$ models. The proposed $d+id$ and $d+ig$ states also conform with experimental evidence that $Sr_2RuO_4$ is a spin singlet, two component order parameter system with a possible TRSB state ~\cite{VanHoveBandfillingTc,KnightShiftPustogow, AlexPetsch,UltrasoundEviTwo,ThermoEvi, GrinenkoSplitTRSB,TRSBSCinSrRuO4,KerrEffect}.

We note that there is a remarkably small discrepancy, of about $0.2$K, between the calculated value of $T_c^{MAX}$ as we move through the topological transition, Fig.~\ref{fig:Tc_all}(d)-(e) compared to experiment~\cite{Tc_plot_not_hicks,SteppkeStrongTc, ResistivityVHS}. This discrepancy
can most likely be attributed to the simplified choices of parameters we made in the tight-binding model 
(for example the neglect of strain induced changes in $t'$ in $V({\bf r}-{\bf r}')$). Alternatively this discrepancy may point  perhaps to something more profound requiring the full three band solution of the gap equation 
or strong correlation effects related to the changed topology of the
full Fermi surface.  


We note that the peak in $T_c$ is roughly similar for both the
pure $d_{x^2-y^2}$ pairing state, the $d+id$ pairing state and for the $d+ig$ pairing state, see Figs.~\ref{fig:Tc_all}~(d)-(e). 
In the above cases the leading factor in the changes in $T_c$ appears to be arising from the approach of the topological VHS to the Fermi energy.
The VHS moves \textit{towards} the Fermi Energy causing an increase in the density of states at the Fermi Energy as shown in Fig.~\ref{fig:dos-all}(a). This corresponds to the maximum of $T_c$ at the Lifshitz point and then a decrease in $T_c$ once the VHS has moved through the Fermi Energy and the effective dos decreases. For the pure $d_{x^2-y^2}$ pairing state
we find that $T_c$ drops well below the original $1.5$K for anisotropies 
corresponding to strains of about $\epsilon_{xx} \sim -1\%$, but this is not seen experimentally. In experiments the $T_c$ appears to level off at about $1.2$K 
for strains of about  $\epsilon_{xx} \sim -1\%$~\cite{Tc_plot_not_hicks}.  This behavior is closer to what we find for the $d+id$ and $d+ig$ cases, shown in Fig.~\ref{fig:Tc_all}(d)-(e). As is clear
from that figure the large strain limit is dominated by $d_{xy}$ or $g$ pairing rather than $d_{x^2-y^2}$
as they cross over at some value of the strain after the Lifshitz transition. Exactly where this occurs will depend on the degeneracy (assumed accidental in our model) leading to the
$d+id$ or $d+ig$ pairing state.  The $d+id$ and $d+ig$ pairing states also correspond to superconducting states 
with TRSB, which would be consistent with $\mu$SR and Kerr effect experiments~\cite{TRSBSCinSrRuO4,KerrEffect}. However, note that the accidental degeneracy
required for these states must somehow be preserved under both isotropic strain and disorder 
~\cite{UnsplitSCandTRSB}. This seems unlikely for an accidental 
degeneracy, but might be possible if there is some higher "hidden" symmetry present beyond that simply required by the $D_{4h}$ lattice space group (for example $d_{x^2-y^2}$ and $d_{xy}$ are degenerate in hexagonal, $D_{6h}$, or cylindrical, $D_{\infty h}$, point groups).

A note of interest for the 'd-wave' models considered here 
is that once the lattice is strained, the d-wave model has an automatic 
anisotropy between $x$ and $y$ directions, which provides an implicit mixing 
of two non degenerate order parameters  $d_{x^2-y^2}$ and $s^\pm$. However the calculations shown in Fig.~\ref{fig:Tc_all}(a)  that this mixing is almost negligible at the strains considered, and so it is still meaningful to consider a ``pure'' $d_{x^2-y^2}$ 
paring symmetry even in the slightly
strained lattice.

Note that the other model we considered here with TRSB the $p_x+ip_y$ state
has a degeneracy required by symmetry and so would not be ruled out by the 
experiments of Grimenko {~\cite{UnsplitSCandTRSB}}.~
The chiral-p dos shows the maximum of the dos for $p_y$ moving towards the Fermi energy and $p_x$ moving away from the Fermi energy, but neither $p_x$ or $p_y$ has a significant peak in the weighted
density of states, Fig.~\ref{fig:dos-all} and so there is no significant uplift in $T_c$
near the Lifshitz point.
Our calculated
$T_c$ splittings in Fig.~\ref{fig:Tc_all}(c) are a poor match to the experiments. Furthermore, they show linear not quadratic change in the higher $T_c$ value with strain, inconsistent with experiment ~\cite{HicksQuadraticStrain}, and secondly the two $T_c$ values would lead to a double
peak in specific heat, which is not seen experimentally~\cite{Tc_plot_not_hicks}. And finally, the T$_c$ past the Lifshitz transition is not in agreement with experiment.
Note that we have not explicitly calculated the different $d+id$ state with 
TRSB symmetry ($d_{xz}+i d_{yz}$) since this requires a three dimensional
pairing model~\cite{ReenaPaper,Suh_2020}  
and cannot be realised in the 2-dimensional Fermi surface model considered here
(there is a symmetry required 
nodal gap in the $k_z=0$ plane).  But based on the symmetry arguments of 
Appendix 1 and the analogy to the $p_x+i p_y$ case we believe that this state would
also be a poor match to experiment, for example with a linear not
quadratic $T_c$ splitting for small strains.

In summary, despite its simplicity, the one band model has accurately reproduced experimental data and provided insight into what happens through the Lifshitz transition. 
In Fig.~\ref{fig:top_trans}~(c) and Fig.~\ref{fig:Fermi-all} we can see the nodal lines on the Fermi surface for each type of pairing symmetry considered here. When the nodal lines match up with the opening of the VHS, we see a decrease in $T_c$ seemingly from the fact that there simply are not enough available electrons to contribute to the superconducting state in the crucial energy range. In contrast, the only pairing states which are expected
to give a major uplift in $T_c$ are either pure or mixed states with a dominant $d_{x^2-y^2}$
component at the maximum $T_c$.  The simple model therefore
 appears as a powerful tool but should be tested further, perhaps combined with density functional theory to model more precisely a full band system in a tetragonal and strained lattice. 
It would also be useful to apply these same techniques to other unconventional superconductors,
since uniaxial strain provides a unique and direct probe of the pairing state symmetry.

\begin{acknowledgments}
This work was supported by the UK Engineering and Physical Sciences Research Council (EPSRC) grant S100154-102 for the University of Bristol, Faculty of Science.
This study was carried out using the computational facilities of the Advanced Computing Research Centre, University of Bristol http://www.bris.ac.uk/acrc/.

\end{acknowledgments}

\appendix

\section{Appendixes}

\subsection{\label{sec:Group} Group Theory}

We can use Ginzburg-Landau theory combined with some considerations from group theory to examine the effects of the lattice symmetry changes induced by uniaxial strain on different symmetry pairing states. In systems with a single superconducting order parameter we expand the free energy density in the usual form
\begin{equation}
 f_s[\psi] =  \dot{\alpha} (T - T_c) |\psi|^2 + \frac{\beta}{2}  |\psi|^4  + \frac{1}{2m} |(-i \hbar \mathbf{\nabla} -2e {\bf A}) \psi|^2
    \end{equation} 
 \cite{schrieffer1999theory}, where $\dot{\alpha}$, $\beta$ and $m$ are positive constants. In the usual singlet BCS pairing case the order parameter
 $\psi$ is proportional to the BCS gap function - $\Delta$. The finite value of $|\psi|$ or $|\Delta|$ appearing below $T_c$ represents the spontaneous breaking of global gauge symmetry in the superconducting phase, while no other crystallographic or spin symmetries are broken
 at $T_c$. 
 
A more general phase transition breaking both gauge and lattice symmetries can be represented by
a Ginzburg-Landau free energy of the form  \cite{JamesGroupTheory}
\begin{eqnarray}
     f_s[\left\{ \psi_i \right\} ] &=&  \alpha_{ij}(T) \psi^*_i \psi_j  + \frac{1}{2} \beta_{ijkl}
     \psi^*_i \psi^*_j \psi_k \psi_l \nonumber \\
     & & + \frac{1}{2} K_{ijkl}  (D_i \psi^*_j ) (D_k \psi_l) 
     \label{eq:GL2}
\end{eqnarray}
where we have a set of order parameter components $\left\{ \psi_i \right\}$,
the covariant derivatives $D_i= (-i \hbar\mathbf{\nabla} -2e{\bf A})$ and 
summation convention of the repeated indices is implied.
The principles of spontaneous symmetry breaking phase transitions imply that the
parameters in the Ginzburg-Landau expansion $\alpha_{ij}$, $\beta_{ijkl}$ and $K_{ijkl}$ possess the full symmetries of the normal state, which in this case implies full lattice point group symmetries.  The unique property of applied uniaxial strain, compared to isotropic pressure, doping or other bulk lattice changes, is that by changing the lattice point group symmetry
it provides direct symmetry specific information on the superconducting phase transition.

To be more specific let us now consider the specific case of a two-dimensional square lattice, as 
considered in the main part of this paper. In Eq.\ref{eq:GL2} the matrix of quadratic terms
$\alpha_{ij}(T)$ can be decomposed into sub-matrices corresponding to the different irreducible
representations of the symmetry group,$D_{2h}$ or $D_{4h}$ in this case. Above $T_c$ all eigenvalues of the matrix are positive and the phase transition $T_c$
is the temperature at which one of the eigenvalues first becomes zero. The corresponding 
eigenvector(s), $\psi_i$, therefore belong to a single irreducible representation of the symmetry group  (assuming no ``accidental'' eigenvalue degeneracies, a possibility discussed further below).  

The character tables for the strained, orthorhombic, $D_{2h}$ and unstrained, tetragonal, 
lattices are given in tables  \ref{table:D2h} and  \ref{table:D4h}. For simplicity we omit the additional parity label, noting that singlet superconductivity implies even parity while triplet pairing implies odd parity, since both crystal structures are centrosymmetric. Considering first the unstrained case the possible order parameter symmetries are derived from table \ref{table:D4h}
and the specific cases we consider in this paper, motivated by proposed pairing states of Sr$_2$RuO$_4$, are given in table \ref{table:gap}. The singlet cases, $s$, extended-$s$ and $d$-wave pairing correspond to a superconducting gap function on the Fermi surface of the form
$\Delta_k$ indicated by the function $\Gamma({\bf k})$ (ignoring multiplicative constants)
and the triplet case represents the chiral triplet order parameter ${\bf d}_{\bf k} \propto 
\Gamma{\bf k} \hat{\bf z}$ ~\cite{ManSigristRice_Sr2RuO4, alexandrov2003theory}).

In order to understand the symmetry induced changes in $T_c$ in these various model pairing states it is helpful to consider the  lattice strain $\epsilon_{ij}$ as an additional parameter in
the Ginzburg-Landau theory.  In two dimensions we have only to consider 
the tensor components $\epsilon_{xx}$, $\epsilon_{yy}$ and $\epsilon_{xy}$. The shear strain 
$\epsilon_{xy}$ can be classified as perturbation with symmetry $B_2$ in Table \ref{table:D4h}, while the uniaxial strains $\epsilon_{xx}$ and $\epsilon_{yy}$ combine both $A_1$ and $B_1$ symmetry components. The pure $A_1$ component is 
\begin{equation}
 \epsilon_{xx} + \epsilon_{yy}
    \end{equation}
while the purely $B_1$ symmetry component is 
    \begin{equation}
        \epsilon_{xx} - \epsilon_{yy} .
\end{equation}
In the uniaxial strain experiments conducted for Sr$_2$RuO$_4$ by Hicks {\it et al.}  \cite{HicksQuadraticStrain} and Steppke {\it et al.}  \cite{SteppkeStrongTc} 
the applied strain includes components of both $A_1$ and $B_1$ types, in proportions
dictated by the Poisson ratio. 
The $A_1$ component of the uniaxial strain does not change the symmetry and so we will not consider it further here (but of course this does not imply that the symmetric component does not affect $T_c$, for example by moving the Fermi surface towards or away from a topological transition and van Hove singularity). 
We will also 
not consider the shear components further in this paper, although note that that this
strain component is discussed in the works of Contreras and Moreno \cite{contreras2019anisotropic} and 
\cite{UltrasoundEviTwo}.   Finally note that the symmetry considerations given here 
are essentially unchanged whether we consider the strain $\epsilon_{ij}$ or stress $\sigma_{ij}$
as the external perturbation.

We focus now only on the symmetry breaking $B_1$ component of the strain $\epsilon_{xx}-\epsilon_{yy}$ as a perturbation in the Ginzburg Landau expansion.
If we condider first a single component order parameter, for example local $s$-wave ($A_1$), 
extended $s$-wave ($A_1$) or pure $d$-wave ($B_1$) pairing states from Tables \ref{table:D4h}
then the leading order term in the the Ginzburg Landau expansion determining $T_c$ 
is of the form
\begin{equation}
    f_s(\psi) =  \dot{\alpha} (T - T_c(0)) |\psi|^2  +  c |\psi|^2 (\epsilon_{xx}-\epsilon_{yy})^2 + \dots .
\end{equation}
This form is dictated by the requirement that the terms in the Ginzburg Landau expansion all have full lattice symmetry ($A_1$) and the group representation identity $B_1 \times B_1 = A_1$ implies
that only a quadratic function of strain is allowed. 
So we immediately see that the $T_c$ changes at most as a quadratic function of the $B_1$ component of strain
\begin{equation}
 T_c (\epsilon) = T_c(0) -  \frac{c}{\dot{\alpha}} (\epsilon_{xx}-\epsilon_{yy})^2 + \dots .
  \end{equation}
Note that the unknown coupling parameter $c$ could be either positive or negative, and so 
symmetry alone does not tell us if $T_c$ increases or decreases, only that the change is a quadratic function of the perturbation.

A more interesting case is provided by the chiral triplet state, $p+ip$ in Table \ref{table:gap}. 
The $E$ symmetry of the square lattice dictates that there are two order parameter components $\psi_x$ and $\psi_y$ with a degenerate $T_c$ in the square lattice. This degeneracy is necessarily broken by uniaxial strain, since the character table \ref{table:D2h} contains no degenerate irreducible representations. To quadratic order the appropriate Ginzburg Landau expansion becomes
\begin{eqnarray}
    f_s(\psi_x,\psi_y) &=&  \dot{\alpha} (T - T_c(0)) \left( |\psi_x|^2 + |\psi_y|^2 \right)
    \nonumber \\
    & & +  c  \left( |\psi_x|^2  - |\psi_y|^2 \right) (\epsilon_{xx}-\epsilon_{yy}) + \dots ,
\end{eqnarray} 
where the specific form is determined by the identity $E \times E = A_1 + A_2 + B_1 + B_2$ in $D_{4h}$.
From this we can immediately see that the degenerate transitions split linearly in the strain
with two transition temperatures
\begin{equation}
    T_c(\epsilon) = T_c(0) \pm  \frac{c}{\dot{\alpha}}  | \epsilon_{xx}-\epsilon_{yy} | .
\end{equation}
The specific state of the order parameter below $T_c$ is determined by the higher order terms in the Ginzburg Landau expansion which we do not consider here. But without any further analysis we can immediately infer that between the two transition temperatures there can be no spontaneous time reversal symmetry breaking. This is because the order parameter is non-degenerate ($A_1$ in the orthorhombic group Table \ref{table:D2h}) while TRSB requires more than a single order parameter component.  This picture is consistent with the numerical results presented in the 
main body of the paper above.

The final cases we consider are states where there are two non-degenerate, or ``accidentally'' degenerate order parameters associated with the unstrained lattice.    First we consider the
case of mixed $d$-wave and extended $s$-wave pairing from table \ref{table:gap}. The two order parameters have $B_1$ ($\psi_d$) and $A_1$ ($\psi_s$) symmetry, and so are not expected to be degenerate at zero strain. However the  $B_1$ symmetry breaking strain couples these linearly
in the form
\begin{eqnarray}
    f_s(\psi_d,\psi_s) &=&  \dot{\alpha_d} (T - T_{d}(0)) |\psi_d|^2 +  \dot{\alpha_s} (T - T_{s}(0)) |\psi_s|^2 \nonumber \\ 
    & & +  c  \left[ \psi_d^*\psi_s +  \psi_s^*\psi_d \right] (\epsilon_{xx}-\epsilon_{yy}) + \dots 
\end{eqnarray}  
(this form is dictated by the identity $B_1 \times A_1 = B_1$, combined with the requirement that the free energy be real valued).
If the two unperturbed $T_c$ values had been degenerate  then again the perturbed $T_c$ values
would again split linearly in the symmetry breaking perturbation $\epsilon_{xx}-\epsilon_{yy}$. But in the numerical model examined in the main body of the paper the two unperturbed $T_c$ values are not close and so the changes in the two transition temperatures are again quadratic in the perturbation, as we see in the numerical results presented above. In fact from the numerical results presented above we can consider the perturbed state a slightly anisotropic form of $d$-wave pairing in which the maximum gaps $|\Delta_{\bf k}|$ in the $k_x$ and $k_y$  lobes of the $d_{x^2-y^2}$ function become slightly unequal
and the gap node, $\Delta_{\bf k}=0$ is slightly shifted away from the line $ak_x=bk_y$ in the orthorhombic Brillouin  zone. 
Finally we consider the cases of a degeneracy or near degeneracy of the form $d+id$ combining $B_1$ and $B_2$ pairing symmetries. 
Denoting the corresponding order parameters $\psi_1$ and $\psi_2$, respectively, we can see that the uniaxial strain has no direct coupling term in the Ginzburg Landau theory (since $B_1 \times B_2 = A_2$ in $D_{4h}$), and so each of the two components couples independently to the uniaxial strain 
\begin{eqnarray}
    f_s(\psi_1,\psi_2) &=&  \dot{\alpha_1} (T - T_{1}(0)) |\psi_1|^2 +  \dot{\alpha_2} (T - T_{2}(0)) |\psi_2|^2 \nonumber \\
    && + (  c_1  |\psi_1|^2 + c_2  |\psi_2|^2 )(\epsilon_{xx}-\epsilon_{yy})^2  + \dots .
    \nonumber \\
    & & \ 
    \label{eq:GL5}
\end{eqnarray}
The same conclusion can be reached by consideration of the character table \ref{table:D2h}. 
The $B_2$ state of $D_{4h}$ is preserved in $D_{2h}$ (it becomes renamed as $B_1$), while the $B_1$ state in $D_{4h}$ becomes $A_1$ in $D_{2h}$, and so these two distinct symmetry states remain distinct despite the change from tetragonal to orthorhombic lattices. This
situation also applies in the case of $d+ig$ pairing\cite{Kivelson_2020}, except that now
the two order superconducting parameters are $B_{1g}$ and $A_{2g}$. These are not coupled
at any order by a strain of $B_{1g}$ symmetry in the quadratic terms of the Ginzburg-Landau theory.
We can conclude from Eq.\ref{eq:GL5} that although the uniaxial strain my be expected to break the ``accidental'' degeneracy in the square lattice for $d+id$ or $d+ig$ type pairing , it does so as at least a quadratic rather than as a linear function of the symmetry breaking perturbation, which agrees with our numerical results presented above.



\begin{table}[h]
\caption{$D_{2h}$ Character Table}
\begin{tabular}{ c c c c c c }
\hline\\
Basis functions & $\Gamma$  & E & $C_2$ & $C_2^\prime$ & $C_2^{''}$ \\
\hline \\
 $x^2,y^2,z^2$ & $A_1$ & 1 & 1 & 1 & 1 \\ 
xy & $B_1$ & 1 & 1 & -1 & -1 \\  
 xz, x &$B_2$ & 1 & -1 & 1 & -1 \\
 yz, y & $B_3$ & 1 & -1 & -1 & 1 \\ 
 \hline\\
\label{table:D2h}

\end{tabular}
\end{table}

\begin{table}[h]
\caption{$D_{4h}$ Character Table}
\begin{tabular}{ c c c c c c c}
\hline\\
Basis functions & $\Gamma$  & E & $2C_4$ & $C_2$ & $2C_2^\prime$ & $2C_2^{''}$\\
\hline\\
$x^2+y^2,z^2$ & $A_1$ & 1 & 1 & 1 & 1 & 1\\ 
$xy(x^2 - y^2)$ & $A_2$ & 1 & 1 & 1 & -1 & -1\\  
$x^2-y^2$ & $B_1$ & 1 & -1 & 1 & 1 & -1\\
 xy & $B_2$ & 1 & -1 & 1 & -1 & 1\\ 
(x,y), (xz,yz) & $E$ & 2 & 0 & -2 & 0 & 0 \\
\hline\\
 \label{table:D4h}
\end{tabular}
\end{table}

\begin{table}[h!]
\caption{Gap Symmetry: The table shows the specific, non-normalized, 
gap function forms considered in our numerical calculations, 
presented for the unperturbed square lattice case, with the 
corresponding irreducible representation as given in the D4h table (\ref{table:D4h}). For convenience we set the lattice constant $a=1$ below.}
\begin{tabular}{ c c c }
\hline \\
Symmetry & $\Gamma({\bf k})$ & Rep.\\
\hline \\
s-wave & 1 & $ A_1$\\ 
$s^{\pm}$& $cos(k_x) + cos(k_y)$ & $A_2$\\
$d_{x^2 - y^2}$ & $ cos(k_x) - cos(k_y)$ & $B_1$\\
$p + ip$ & $sin(k_x) + isin(k_y)$ & $E$\\
$d + id$ & $[cos(k_x)-cos(k_y)] + i sin(k_x)sin(k_y) $ & $B_1$ + $B_2$\\
$d + ig$ & $[cos(k_x)-cos(k_y)](1 + i sin(k_x)sin(k_y)) $ & $B_1$ + $A_2$\\
\hline\\
\label{table:gap}
\end{tabular}
\end{table}

\newpage



\end{document}